\documentclass[
letter,
twocolumn,
superscriptaddress,
amsmath,
amssymb,
prl,
nofootinbib,
showkeys,
10pt,
floatfix
]{revtex4-1}

\usepackage{graphicx}
\usepackage{dcolumn}
\usepackage{bm}
\usepackage{xcolor}
\usepackage{ulem}
\usepackage{hyperref}
\usepackage{natbib,hyperref}
\usepackage{cleveref}
\usepackage{amsmath}
\usepackage{float}
\usepackage{svg}
\usepackage{amssymb}
\usepackage{scalerel}
\usepackage{caption}
\usepackage{setspace}
\usepackage{multirow}
\usepackage{booktabs}
\usepackage{siunitx}
\usepackage[misc]{ifsym}
\usepackage{makecell}
\usepackage{array}
\usepackage{CJK}

\graphicspath{{figures/}} % 定义所有的图片文件在 figures 子目录下

\makeatletter
\pretocmd\frontmatter@keys@format{\addvspace{20\p@}}{}{}
\makeatother

\begin{document}

\title{
Quafu-RL: The Cloud Quantum Computers based Quantum Reinforcement Learning
}

\author{BAQIS Quafu Group}
%\affiliation{Beijing Academy of Quantum Information Sciences, Beijing 100193, China}
%\affiliation{School of Mathematical Sciences, Nankai University, Tianjin, 300071, China}

% \author{Hong-Ze Xu}
% \affiliation{Beijing Academy of Quantum Information Sciences, Beijing 100193, China}

% \author{Meng-Jun Hu \Letter} \email{humj@baqis.ac.cn}
% \affiliation{Beijing Academy of Quantum Information Sciences, Beijing 100193, China}

% \date{}

\begin{abstract}
With the rapid advancement of quantum computing, hybrid quantum-classical machine learning has shown numerous potential applications at the current stage, with expectations of being achievable in the noisy intermediate-scale quantum (NISQ) era. Quantum reinforcement learning, as an indispensable study, has recently demonstrated its ability to solve standard benchmark environments with formally provable theoretical advantages over classical counterparts. However, despite the progress of quantum processors and the emergence of quantum computing clouds, the implementation of quantum reinforcement learning algorithms utilizing parameterized quantum circuits (PQCs) on NISQ devices remains infrequent. In this work, we take the first step towards executing benchmark quantum reinforcement problems on real devices equipped with at most 136 qubits on BAQIS Quafu quantum computing cloud. The experimental results demonstrate that the policy agents can successfully accomplish objectives under modified conditions in both the training and inference phases. Moreover, we design hardware-efficient PQC architectures in the quantum model using a multi-objective evolutionary algorithm and develop a learning algorithm that is adaptable to quantum devices. We hope that the Quafu-RL be a guiding example to show how to realize machine learning task by taking advantage of quantum computers on the quantum cloud platform.\par



\end{abstract}

\maketitle

%\onecolumngrid

\section{Introduction}\label{sec1}
Recent years have witnessed a huge development of quantum computing, encompassing quantum hardware and abundant software platforms. Despite the proposition of quantum supremacy \cite{bib1, bib2, bib3}, limitations persist in the current noisy intermediate-scale quantum (NISQ) regime, primarily stemming from a constrained number of qubits, quantum state decoherence, and imprecise quantum operations \cite{bib4, bib5}. With the obstacle of constructing ideal quantum processing units (QPUs) in the short term, a current trend indicates that it is necessary to build hybrid quantum-classical infrastructure to explore useful applications while making efforts to scale up QPUs \cite{bib6}. Quantum computing clouds such as IBMQ, Amazon Braket and Quafu \cite{bib40} serve as bridges to connect developers in quantum software applications with advanced quantum computers in NISQ era. Although quantum computing is promising to solve some intractable classical NP-hard problems, deficient quantum devices nowadays favorably excel at hybrid algorithms leveraging both quantum circuits and classical computing.\par

Variational quantum algorithms (VQAs) that run parameterized quantum circuits (PQCs) on quantum devices while integrating classical optimizer for parameter optimization satisfy the desire to demonstrate applications on NISQ devices \cite{bib7} and have been wildly explored. Researchers in VQAs mainly focus on quantum chemistry \cite{bib8,bib9,bib10}, quantum optimization \cite{bib11,bib12,bib13} and quantum machine learning (QML) \cite{bib14,bib15} including
interesting problems such as classification \cite{bib16,bib17,bib18,bib19}, generative adversarial networks \cite{bib20,bib21,bib22,bib23} and related theoretic analysis \cite{bib24,bib25,bib26}.\par 

Reinforcement learning (RL) \cite{bib27}, as an essential area in modern machine learning research, receives comparatively later attention in VQA-based approaches \cite{bib28,bib29,bib30,bib31}. Nonetheless, it is hard for these proposed PQC agents to reach satisfactory performance in benchmark environments from OpenAI Gym \cite{bib32}. Until recently, policy-based \cite{bib33} and value-based \cite{bib34} quantum reinforcement learning make breakthroughs in both solving standard benchmark tasks and theoretical learning advantage over classical algorithms. In \cite{bib33,bib34}, apart from hyperparameters tuning, the importance of PQC architectures in successful RL training and better performance is proposed. This issue inspires some evolutionary methods \cite{bib35,bib36,bib37,bib47,bib48,bib49} to search suitable PQC architectures without human ingenuity in the aim of balancing expressivity \cite{bib38, bib39} and trainability.\par

In this study, we leverage the theoretical guarantees and experimental validation of parameterized quantum policies for reinforcement learning, along with the availability of quantum computing clouds like Quafu, which offer scalable and stable services over the long term. With these resources, we are able to apply a benchmark RL task, CartPole, to real quantum devices and conduct numerous experiments. Firstly, we utilize a multi-objective evolutionary approach to search for the most suitable PQCs that can construct policies for higher performance as well as lower entanglement and design a learning algorithm for quantum devices. Next, we select the best policy and execute it on available quantum resources that include 10-qubit, 18-qubit, and 136-qubit quantum computers. Our experiments demonstrate that the environment can be successfully solved with some relaxation of the original goals. Overall, our proposed methods and pipelines offer a perspective for conducting PQC-based RL policies and other QML experiments on NISQ devices.\par

    \begin{figure*}
      \centering
      \includegraphics[width=16cm]{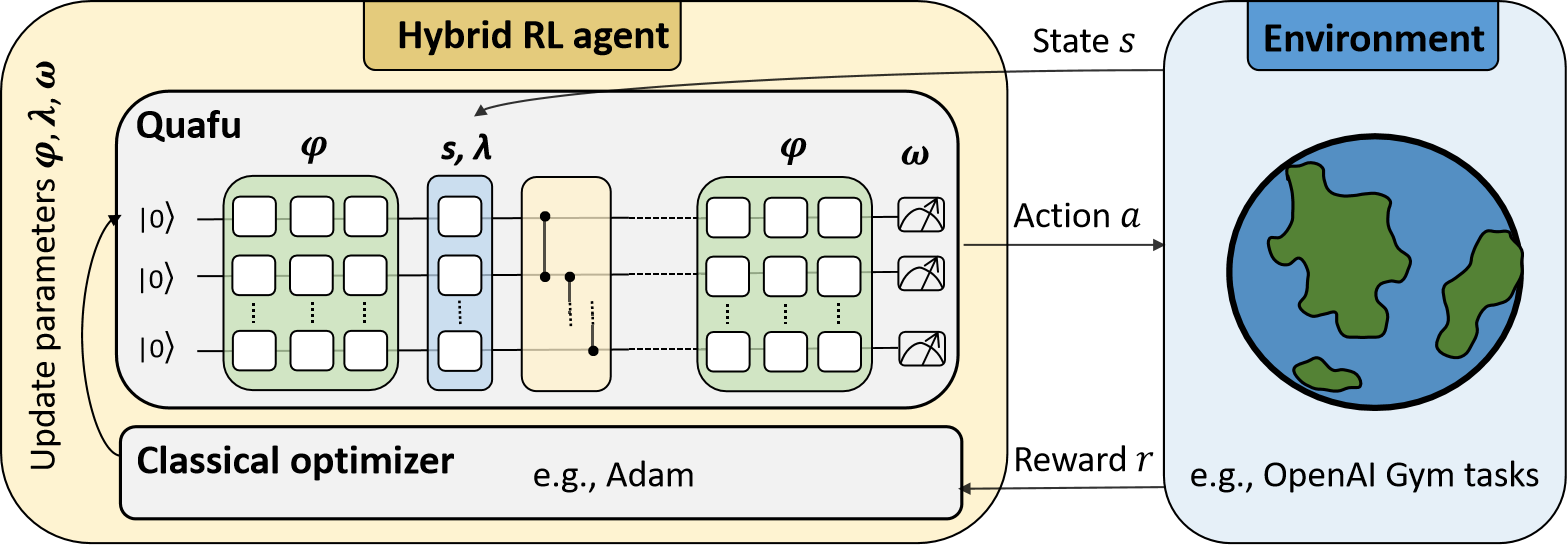}
      \caption{\textbf{Quantum reinforcement learning with Quafu.} The entire pipeline can be described as follows: the RL agent, which is composed of a hardware-efficient PQC designed specifically for this task, receives a state $s$ from the environment. The agent is then evaluated on Quafu and generates a policy $\pi_\theta\left(a \mid s\right)$ to sample an action $a$ and receive a feedback reward $r$. The parameters within the PQC are updated using a classical optimizer and training algorithm introduced later.}
      \label{fig1}
   \end{figure*}

\section{Reinforcement learning}\label{sec2}
Reinforcement learning illustrates a problem that an agent learns to make best decisions to get maximum cumulative rewards through interactions with the environment \cite{bib27, bib41}. Explicitly speaking, as shown in FIG. \ref{fig1}, an agent first observes a state $s \in \mathcal{S}$ from the environment. Then, according to the policy $\pi$, which can be some deterministic tables or neural networks with trainable parameters, the agent takes an action $a$ from all possible actions $\mathcal{A}$. Additionally, a policy can be further described as a mapping from states to the probability of performing each possible action $\pi(a \mid s)$. After executing the action $a$, the environment returns a new state $s^{\prime}$ and a reward $r$. Aforementioned transitions usually formulated as $p\left(s^{\prime} \mid s, a\right)$ describe the dynamics of the environment. In order to formalize the RL objective, discount factor $\gamma$ is introduced which lies in $[0,1]$. And the corresponding discounted return is defined as the weighted sum of future rewards $G_t=\sum_{k=0}^{T-t-1} \gamma^k r_{t+k+1}$, which indicates the rewards obtained starting at time $t$ until the final time $T$. In essence, gathering above all elements, reinforcement learning is mathematically modeled with Markov Decision Process (MDP) as a tuple $(\mathcal{S}, \mathcal{A}, p, G, \gamma)$.\par

\subsection{Policy gradient methods}
Reinforcement learning algorithms can be classified into two main categories: value-based and policy-based methods. Value-based methods aim to find the policy that maximizes the value function, while policy-based methods directly attempt to find the optimal policy by proposing a parameterized $\pi_{\boldsymbol{\theta}}(a \mid s)$ and optimizing its parameters $\boldsymbol{\theta}$. In this paper, we mainly focus on the policy gradient methods.\par
In most cases, policy gradient methods compute the gradient of the expected total return $\nabla_{\boldsymbol{\theta}} \mathbb{E}\left[G \mid \pi_{\boldsymbol{\theta}}\right]$ and apply gradient ascent to update parameters. Specifically, the expectation value is taken over the trajectories $\tau$ which collect all states, actions and rewards $s_0, a_0, r_1, s_1, a_1, r_2, s_2, a_2, \ldots, s_T$ within one episode. Then, consistent with policy gradient theorem in \cite{bib42}, the expectation can be rewritten as
    \begin{equation}
        \nabla_{\boldsymbol{\theta}} \mathbb{E}\left[G \mid \pi_{\boldsymbol{\theta}}\right] =\mathbb{E}\left[G \sum_{t=0}^{T-1} \nabla_{\boldsymbol{\theta}} \log \pi_{\boldsymbol{\theta}}\left(a_t \mid s_t\right) \mid \pi_{\boldsymbol{\theta}}\right].
        \label{eq1}
    \end{equation}

\section{PQC-based quantum reinforcement learning}\label{sec3}
\begin{figure*}
      \centering
      \includegraphics[width=16cm]{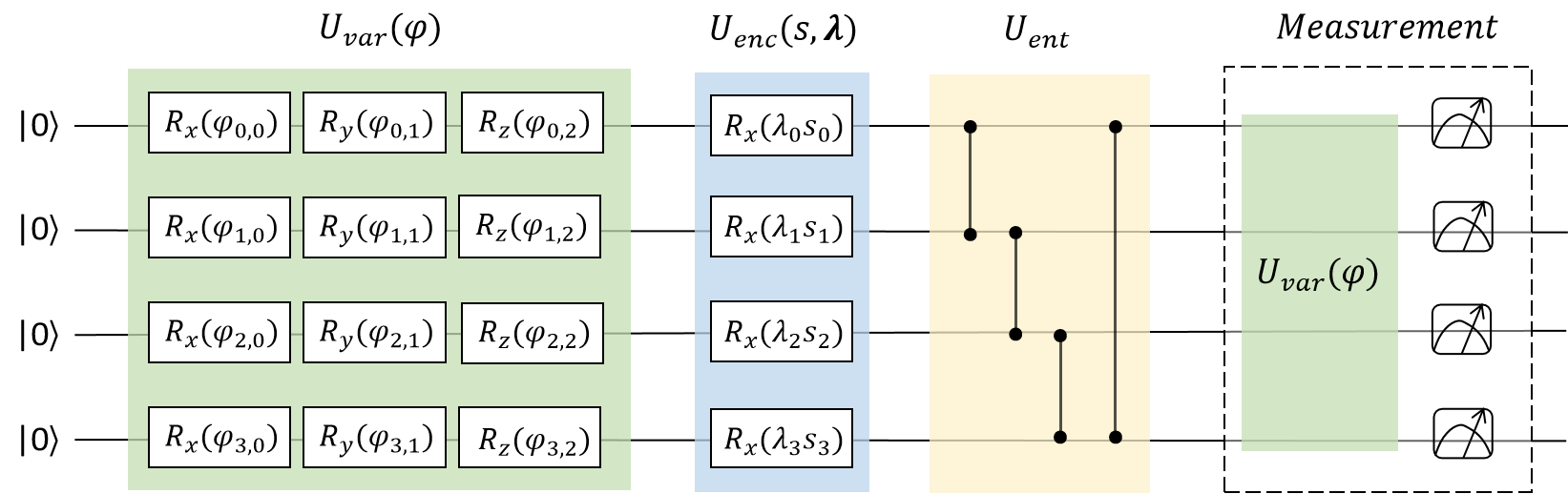}
      \caption{\textbf{PQC architecture components.} The example 4-qubit circuit consists of four basic PQC architecture components including variational PQC denoted as $U_{v a r}(\varphi)$ with trainable parameters $\varphi$, data-encoding PQC termed as $U_{e n c}(s, \lambda)$ with input state $s$ and scaling parameters $\lambda$, entanglement identified as $U_{\text {ent }}$ and the final measurement part.}
      \label{fig2}
   \end{figure*}

In the quantum realm, parameterized quantum circuits which can be represented by a unitary $U(s, \boldsymbol{\theta})$ with an input state $s$ and trainable parameters $\boldsymbol{\theta}$ are utilized to build reinforcement learning policies, analogous to the function of classical neural networks. Recent studies \cite{bib15,bib43,bib38,bib39} have revealed some hardware-efficient PQCs, typically composed of variational PQCs, data-encoding PQCs, entanglements and a final measurement. In the context, as displayed in FIG. \ref{fig2}, a variational PQC ($U_{v a r}(\varphi)$) refers to a circuit made up of single-qubit rotations $R_x, R_y, R_z$ acting on each qubit, assigning rotation angles $\varphi$ as trainable parameters. Moreover, a data-encoding PQC ($U_{e n c}(s, \lambda)$) is formed by $R_x$ applied to each qubit, with a state vector $s=\left(s_0, s_{1}, \ldots \right)$ scaled by trainable parameters $\lambda$. Also, an entanglement ($U_{\text {ent }}$) achieves circular entanglement by using multiple controlled-Z gates. At the end, measurement is performed just after a variational PQC.\par

\subsection{Parameterized quantum policies}
The four basic operations described above are the building blocks of the most common PQC architectures. However, in order to solve RL environments and obtain good performances, some data-encoding techniques and readout strategies are crucial for success. To enhance the expressive power, data-reuploading \cite{bib39} which means using data-encoding PQC repeatedly in a circuit is a widely-adopted skill. In \cite{bib33,bib34}, a variational PQC, an entanglement and a data-encoding PQC are assembled as an alternating layer and are recurrently applied to create highly-expressive parameterized quantum policies. Meanwhile, to handle the range of the output of measurements, weighted observables associated to action $a$ are utilized with trainable parameters $\omega$. The expectation value of weighted observables then can be formulated as 
    \begin{equation}
        \left\langle O_a\right\rangle_{s, \boldsymbol{\theta}}=\left\langle 0^{\otimes n}\left|U(s, \varphi, \lambda)^{\dagger} O_a U(s, \varphi, \lambda)\right| 0^{\otimes n}\right\rangle \cdot \omega_{a},
        \label{eq2}
    \end{equation}
considering a n-qubit PQC with input state $s$, rotation angles $\varphi$, scaling parameters $\lambda$ and observable operators $O_a$, where $\boldsymbol{\theta} = (\varphi, \lambda, \omega)$. Furthermore, non-linear activation function softmax is applied on the expectation values $\left\langle O_a\right\rangle_{s, \boldsymbol{\theta}}$, which defines a SOFTMAX-PQC policy as
    \begin{equation}
        \pi_{\boldsymbol{\theta}}(a \mid s)=\frac{e^{\beta\left\langle O_{a}\right\rangle_{s, \boldsymbol{\theta}}}}{\sum_{a^{\prime}} e^{\beta\left\langle O_{a^{\prime}}\right\rangle_{s, \boldsymbol{\theta}}}},
        \label{eq3}
    \end{equation}
where $\beta$ is an inverse-temperature parameter.\par

\subsection{Quantum architecture search}
It is effective to adopt pre-selected alternating layers to construct the quantum policy architecture as shown in the work \cite{bib33}, but a recent work \cite{bib35} proposes an evolutionary quantum architecture search method to build a more flexible PQC policy using the four fundamental architecture components in FIG. \ref{fig2}, achieving higher RL performance with lower depths.\par
To be more precise, the four elementary PQC structures, namely $U_{v a r}(\varphi)$, $U_{e n c}(s, \lambda)$, $U_{\text {ent }}$, and measurement, are represented as genes and encoded into integer spaces with the values 1, 2, 3, and 0, respectively. Afterwards, Non-Dominated Sorting Genetic Algorithm II (NSGA-II) \cite{bib44} is implemented to optimize the quantum architecture search process which iteratively generates a population of candidates through genetic operations like crossover and mutation on the given parents, and selects parents for the next generation based on fitness evaluation with the RL average collected rewards as the objective.\par

\section{Quantum reinforcement learning with Quafu}\label{sec4}
With the assistance of the SOFTMAX-PQC policy and evolutionary quantum architecture search method outlined in the preceding section, we are able to devise a PQC that is both efficient in hardware usage and effective in implementing quantum reinforcement learning tasks on the Quafu cloud platform.\par
    \begin{figure*}
      \centering
      \includegraphics[scale=0.53]{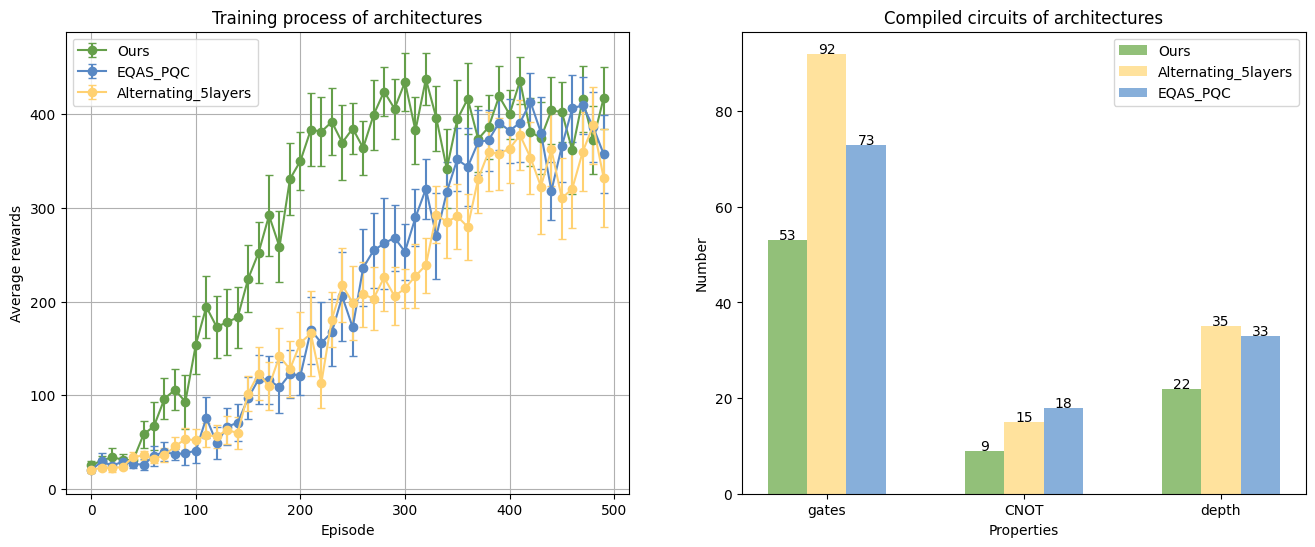}
      \caption{\textbf{Architecture comparison.} Our proposed architecture, 5-alternating-layer method, and EQAS-PQC are compared based on their training performance and compiled circuit properties. The left subplot displays the average performance based on 10 trails, with error bars representing 0.3 times the standard deviation. And the right subplot shows the number of gates, CNOTs, and circuit depth of the compiled circuits.}
      \label{fig3}
   \end{figure*}

\subsection{Architecture selection}
To select a hardware-efficient PQC for executing RL agents on Quafu, we utilize the method of evolutionary quantum architecture search (EQAS-PQC) \cite{bib35}, but take a further step in refining our objective. We design a multi-objective function that not only takes into account RL training performance, but also considers the number of entanglement components $U_{\text {ent }}$. The optimal architecture discovered during the search process for the CartPole environment can be represented using genes as follows: 3 - 1 - 1 - 2 - 1 - 1 - 3 - 1 - 3 - 2 - 1 - 2 - 0. To evaluate the ability and efficiency of our selected architecture, we compare it with the best architecture proposed in \cite{bib35} and the alternating scheme presented in \cite{bib33}.\par
As illustrated in FIG. \ref{fig3}, the left subplot demonstrates that our PQC architecture achieves the best performance and is the first to reach the goal, whereas both the 5-alternating-layer method and EQAS-PQC slowly improve to solve the environment within 500 training episodes. Moreover, the Quafu cloud platform provides a compilation scheme for quantum circuits prior to their execution on the quantum computer. Therefore, certain hardware-sensitive characteristics (such as the number of total gates, the number of two-qubit gate CNOT and the depth of the circuit) of the compiled circuits for the architectures are crucial for achieving optimal performance on Quafu. As shown in the right subplot of FIG. \ref{fig3}, our proposed architecture exhibits the least number of gates and CNOTs, as well as the shortest depth of the compiled circuit. In contrast, the other two methods have significantly more gates, CNOTs, and deeper circuit depths, which clearly indicates their lower hardware efficiency. Some implementation details to note in conducting the experiments shown in FIG. \ref{fig3} are that we have removed the non-nearest neighbor CNOTs from the entanglement part. This is because including these CNOTs would drastically increase the number of gates and circuit depth, without adding significant benefits to the overall success of the task.\par
Briefly, our proposed multi-objective evolutionary quantum architecture search method enables us to identify a suitable PQC architecture that exhibits proficient RL performance and is also hardware-adaptive to the Quafu platform, as described in detail above.\par

\subsection{Learning algorithm}

\begin{tabular}{l}
\hline \textbf{Algorithm 1} REINFORCE with Quafu\\
\hline \textbf{Input:} initialized SOFTMAX-PQC policy $\pi_{\boldsymbol{\theta}}(a \mid s)$, \\
learning rate $\eta$, number of trajectories $N$, maximum time $T$ \\
\textbf{while} True do \\
\quad \textbf{for} $i=1$ to $N$ do \\
\quad\quad Initialize $s_0$ \\
\quad\quad \textbf{for} $t=0$ to $T-1$ do \\
\quad\quad\quad Execute PQC$_{\theta}$ on Quafu and get policy $\pi_\theta\left(a_t \mid s_t\right)$ \\
\quad\quad\quad Take action $a_t \sim \pi_\theta\left(a_t \mid s_t\right)$ \\
\quad\quad\quad Move to next state $s_{t+1}$ and store reward $r_{t+1}$ \\
\quad\quad \textbf{end for} \\
\quad\quad Execute PQC$_{\theta}$ on simulator and get policy $\widehat{\pi}_\theta\left(a_t \mid s_t\right)$ \\
\quad\quad$G^{(i)} \leftarrow \sum_t \gamma^t r_{t+1}$ \\
\quad\quad Compute $\epsilon = \pi_\theta\left(a_t \mid s_t\right) - \widehat{\pi}_\theta\left(a_t \mid s_t\right)$ \\
\quad\quad$z^{(i)} \leftarrow \sum_t \nabla_{\boldsymbol{\theta}} \log \left(\widehat{\pi}_\theta\left(a_t \mid s_t\right) + \epsilon\right)$ \\
\quad\textbf{end for} \\
\quad $\Delta \theta \leftarrow(1 / N) \sum_i G^{(i)} z^{(i)}$ \\
\quad $\boldsymbol{\theta} \leftarrow \boldsymbol{\theta}+\eta \Delta \boldsymbol{\theta}$ \\
\textbf{end while} \\
\hline
\end{tabular} 
\\ \hspace*{\fill} \\

\begin{table*}
  \centering
    \newcolumntype{C}{@{\extracolsep{150em}}c@{\extracolsep{0pt}}}
    \begin{tabular}{cccccccc}
    \hline Environment & Qubits & Actions & Reward & Learning rates & $\gamma$ & $\beta$ & Observables \\
    \hline \makecell{\\[10pt]} CartPole-v1 & 4 & 2 & +1 & $0.01,0.1,0.1$ & 1.0 & 1.0 & {$\left[Z_0 Z_1 Z_2 Z_3\right]$} \\
    \hline
    \end{tabular}
    \caption{\textbf{Hyperparameters of the RL environment.} Learning rates correspond to the parameters $\varphi, \omega, \lambda$.}
  \label{tab1}
\end{table*}

    \begin{figure*}
      \centering
      \includegraphics[width=18cm, height=5.5cm]{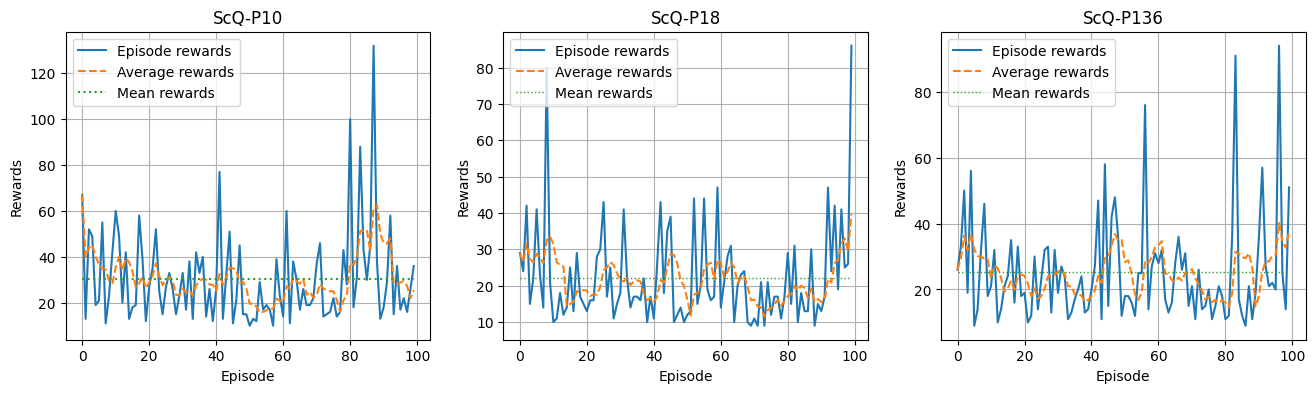}
      \caption{\textbf{RL training performance with Quafu within 100 episodes.} RL agents have been trained on ScQ-P10, ScQ-P18 and ScQ-P136. The blue lines indicate episode rewards within 100 training episodes, the orange dashed lines take a moving average over episode rewards with a window of 5, and the dotted green lines show the mean rewards over all training process.}
      \label{fig4}
   \end{figure*}

After confirming the architecture used in SOFTMAX-PQC policy, it is important to specify an algorithm that connect classical optimizers and the Quafu platform. In \cite{bib33,bib35}, the classical REINFORCE algorithm\cite{bib45} is implemented with quantum simulator. However, with the involvement of quantum devices, there will be slight modifications as shown in Algorithm 1.\par

Combining the selected hardware-efficient PQC and specialized learning algorithm, we have built the whole pipeline of quantum reinforcement learning with Quafu, as pictured in FIG. \ref{fig1}. \par

\section{Numerical results}\label{sec5}
In this section, we investigate a classical benchmark environment called CartPole from the OpenAI Gym \cite{bib32} and apply our proposed PQC policy and algorithm to train the agent on three different quantum devices from Quafu \cite{bib40}: ScQ-P10, ScQP18, and ScQ-P136, which have 10-qubit, 18-qubit, and 136-qubit capabilities, respectively.\par

\subsection{Experimental setups}
We determine the hyperparameters, such as learning rates and observables, based on the standard practices outlined in \cite{bib33}, which are also summarized in TABLE~\ref{tab1}. Furthermore, instead of pursuing the initial goal of typically achieving 500 rewards within 500 episodes in quantum RL, we set a target of obtaining 100 rewards within 100 episodes, considering the high computational cost of training on real quantum computers. Besides, classical optimization process is simulated on TensorFlow Quantum \cite{bib46}.

\subsection{Results}
As presented in FIG. \ref{fig4}, we conduct a series of experiments on the Quafu cloud platform. In the left column of FIG. \ref{fig4}, a RL agent have been trained on ScQ-P10 device and get a maximum reward of 132 within 100 training episodes. Even in noiseless settings as demonstrated in FIG. \ref{fig3}, it is not easy to achieve such objectives with only 100 training episodes. Moreover, observing the rewards averaged by a time window of 5 episodes, the agent is able to reach a stable reward of over 50, which lasts for a certain period of time after 80 training episodes. Additionally, the mean reward over 100 training episodes is approximately 30, which is higher than 20, the reward achieved by random choice. The middle and right subplots in FIG. \ref{fig4} depict the training process of RL agents on ScQ-P18 and ScQ-P136, which accomplish maximum rewards of 86 and 94, respectively. Due to the requirement of only a 4-qubit circuit for our task, having more qubits in a quantum computer does not necessarily lead to superior performance.\par

\begin{figure}[ht]
\centering
\includegraphics[width=0.95\columnwidth]{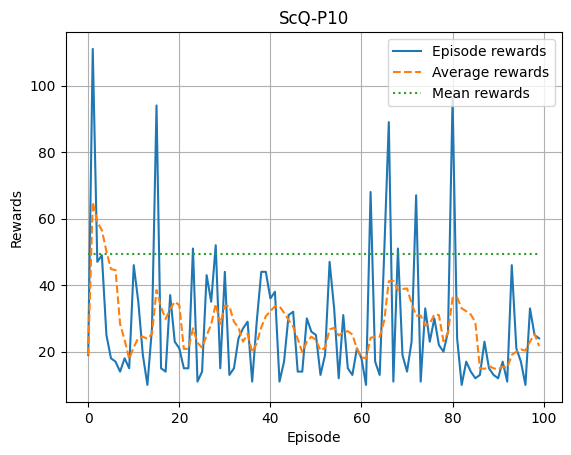}
\caption{\textbf{Inference on ScQ-P10 within 100 episodes.} The interpretation of the lines is similar to that in Figure 4.}
\label{fig5}
\end{figure}

We also evaluate the pre-trained model on ScQ-P10 within 100 episodes as shown in FIG. \ref{fig5}. The results indicate that the RL agent trained on Quafu is capable of attaining a goal of over 100 rewards in the inference stage, with an overall mean reward of approximately 50. These numerical signs provide evidence of the agent's successful training on noisy devices. Although we limit the training and inference to 100 episodes in this study due to the expensive time cost, we believe that extending the training duration will finally lead to improved performance.\par

\section{Conclusion}\label{sec6}
In this study, we implement quantum reinforcement learning on ScQ-P10, ScQ-P18, and ScQ-P136 from the Quafu cloud platform. We diligently choose our hardware-efficient PQCs using a multi-objective evolutionary algorithm and refine the REINFORCE learning algorithm to ensure the feasibility of training RL agents. Experimental results show that RL agents can be successfully trained and evaluated with some relaxation of the original goals within 100 episodes. Based on the experimental results, the potential advantage of quantum reinforcement learning primarily stems from having fewer parameters while maintaining comparable performance.\par
Moving forward, there is potential to train agents on quantum devices with higher objectives, longer episode lengths, and a wider range of test environments. Additionally, the PQC search process could be conducted on real devices under finer restricted conditions to find a more suitable architecture for a specific device. Finally, other algorithms such as PPO could be introduced to stabilize the training process on quantum computers.\par

\par

{\bf Code available:} The corresponding software can be found in: https://github.com/enchanted123/quantum-RL-with-quafu.

\section{Acknowledgments}
This work is supported by the Beijing Academy of Quantum Information Sciences.

% \clearpage

%Bibliography
% \bibliographystyle{unsrt} 
% \bibstyle{plain}
\normalem
\bibliographystyle{iopart-num.bst}
\bibliography{references}  

\begin{CJK*}{UTF8}{gbsn}

\newpage
\onecolumngrid

\vspace{1em}
\begin{flushleft}
{\small BAQIS Quafu Group}

\bigskip
{\small
\renewcommand{\author}[2]{#1$^\textrm{\scriptsize #2}$}
\renewcommand{\affiliation}[2]{$^\textrm{\scriptsize #1}$ #2 \\}

\newcommand{\corrauthora}[2]{#1$^{\textrm{\scriptsize #2}, \ddagger}$}
\newcommand{\corrauthorb}[2]{#1$^{\textrm{\scriptsize #2}, \mathsection}$}

\newcommand{\xBAQIS}{\affiliation{1}{Beijing Academy of Quantum Information Sciences, Beijing 100193, China}}

\newcommand{\xTsinghua}{\affiliation{2}{State Key Laboratory of Low Dimensional Quantum Physics, Department of Physics, Tsinghua University, Beijing, 100084, China}}

\newcommand{\xCAS}{\affiliation{3}{Institute of Physics, Chinese Academy of Sciences, Beijing 100190, China}}

\newcommand{\xFrontier}{\affiliation{4}{Frontier Science Center for Quantum Information, Beijing 100184, China}}

\newcommand{\xCASE}{\affiliation{5}{School of Physical Sciences, University of Chinese Academy of Sciences, Beijing 100190, China}}

\newcommand{\xCASF}{\affiliation{6}{CAS Center for Excellence in Topological Quantum Computation, UCAS, Beijing 100190, China}}

\newcommand{\xNK}{\affiliation{7}{School of Mathematical Sciences, Nankai University, Tianjin, 300071, China}}

%\author{Dong E. Liu}
%\affiliation{Beijing Academy of Quantum Information Sciences, Beijing 100193, China}
%\affiliation{State Key Laboratory of Low Dimensional Quantum Physics,
%Department of Physics, Tsinghua University, Beijing, 100084, China}
%\affiliation{Frontier Science Center for Quantum Information, Beijing 100184, China}

%\affiliation{Beijing Academy of Quantum Information Sciences, Beijing 100193, China}
%\affiliation{Institute of Physics, Chinese Academy of Sciences, Beijing 100190, China}
%\affiliation{School of Physical Sciences, University of Chinese Academy of Sciences, Beijing 100190, China}
%\affiliation{CAS Center for Excellence in Topological Quantum Computation, UCAS, Beijing 100190, China}
%\affiliation{Songshan Lake Materials Laboratory, Dongguan 523808, China}

% Make sure the numbers match up ^^ vv
\newcommand{\BAQIS}{1}
\newcommand{\Tsinghua}{2}
\newcommand{\CAS}{3}
\newcommand{\Frontier}{4}
\newcommand{\CASE}{5}
\newcommand{\CASF}{6}
\newcommand{\NK}{7}

\corrauthora{Yu-Xin Jin (\CJKfamily{gbsn}靳羽欣)}{\BAQIS, \!\NK}
\corrauthora{Hong-Ze Xu (\CJKfamily{gbsn}许宏泽)}{\BAQIS},
\author{Zheng-An Wang (\CJKfamily{gbsn}王正安)}{\BAQIS},
\author{Wei-Feng Zhuang (\CJKfamily{gbsn}庄伟峰)}{\BAQIS},
\author{Kai-Xuan Huang (\CJKfamily{gbsn}黄凯旋)}{\BAQIS},
\author{Yun-Hao Shi (\CJKfamily{gbsn}时运豪)}{\CAS, \!\CASE, \!\CASF},
\author{Bao-Guo Ma (\CJKfamily{gbsn}马保国)}{\CAS, \!\CASE, \!\CASF},
\author{Tian-Ming Li (\CJKfamily{gbsn}李天铭)}{\CAS, \!\CASE, \!\CASF},
\author{Chi-Tong Chen (\CJKfamily{gbsn}陈驰通)}{\CAS, \!\CASE, \!\CASF},
\author{Kai Xu (\CJKfamily{gbsn}许凯)}{\CAS, \!\BAQIS}
\author{Yu-Long Feng (\CJKfamily{gbsn}冯玉龙)}{\BAQIS},
\author{Pei-Liu (\CJKfamily{gbsn}刘培)}{\BAQIS},
\author{Mo Chen (\CJKfamily{gbsn}陈墨)}{\BAQIS}
\author{Shang-Shu Li (\CJKfamily{gbsn}李尚书)}{\CAS, \!\CASE, \!\CASF},
\author{Zhi-Peng Yang (\CJKfamily{gbsn}杨智鹏)}{\BAQIS},
\author{Chen Qian (\CJKfamily{gbsn}钱辰)}{\BAQIS},
\author{Yun-Heng Ma (\CJKfamily{gbsn}马运恒)}{\BAQIS},
\author{Xiao Xiao (\CJKfamily{gbsn}肖骁)}{\BAQIS},
\author{Peng Qian (\CJKfamily{gbsn}钱鹏)}{\BAQIS},
\author{Yanwu Gu (\CJKfamily{gbsn}顾炎武)}{\BAQIS},
\author{Xu-Dan Chai (\CJKfamily{gbsn}柴绪丹)}{\BAQIS}
\author{Ya-Nan Pu (\CJKfamily{gbsn}普亚南)}{\BAQIS},
\author{Yi-Peng Zhang (\CJKfamily{gbsn}张翼鹏)}{\BAQIS},
\author{Shi-Jie Wei (\CJKfamily{gbsn}魏世杰)}{\BAQIS},
\author{Jin-Feng Zeng (\CJKfamily{gbsn}曾进峰)}{\BAQIS},
\author{Hang Li (\CJKfamily{gbsn}李行)}{\BAQIS},
\author{Gui-Lu Long (\CJKfamily{gbsn}龙桂鲁)}{\Tsinghua, \!\BAQIS},
\author{Yirong Jin (\CJKfamily{gbsn}金贻荣)}{\BAQIS},
\author{Haifeng Yu (\CJKfamily{gbsn}于海峰)}{\BAQIS},
\author{Heng Fan (\CJKfamily{gbsn}范桁)}{\CAS, \!\BAQIS, \!\CASE, \!\CASF},
\author{Dong E. Liu (\CJKfamily{gbsn}刘东)}{\Tsinghua, \!\BAQIS, \!\Frontier},
\corrauthorb{Meng-Jun Hu (\CJKfamily{gbsn}胡孟军)}{\BAQIS},

\bigskip

\xBAQIS
\xTsinghua
\xCAS
\xFrontier
\xCASE
\xCASF
\xNK

{${}^\ddagger$ These authors contributed equally to this work.}\\
{${}^\mathsection$ Corresponding author: humj@baqis.ac.cn}

}
\end{flushleft}

% \section{Supporting Information}
% Supporting Information is available 

\end{CJK*}
\end{document}